# Formation of the Moon: a new mechanism

Mukesh Gupta[*]

**Abstract**

Understanding the Moon's formation mechanism is necessary for studying not only the Moon itself, but also the evolution, formation, habitability, and structure of other planets and the moons in the Solar system and in extrasolar planetary systems. In this paper, I suggest a mechanism of the Moon formation. I posit that the Moon came into existence from the same gaseous cloud that the Earth originated from. The paper concludes that the moons can only form at the time of planet formation in a parallel and simultaneous process, and the new moons cannot form in a grown up Solar system. The Earth-Moon system is not a special exception in nature; rather it evolved following the same laws that other planets/moons have evolved from. This perspective successfully elucidates what happened between the disk formation and the accumulation of the Moon from the disk.

## 1. Introduction

Even after decades of research, planetary scientists are still ambiguous of how the Moon came into existence (Canup, 2012; Elkins-Tanton, 2013). The latest speculation is that the Moon came into existence because of a giant impact with the proto-Earth. Hartmann and Davis (1975) originally proposed collision of a large body with the Earth that possibly ejected iron-deficient crust, forming cloud of volatile dust that could form the Moon. The subsequent cooling of the splattered matter resulted in the Moon we see today. However, this hypothesis could not account for several observations to be discussed in the next section. The fission from the Earth triggered by the Earth core formation is the most probable origin of the Moon (Wise, 1969). Another scenario proposed that the fast spinning Earth became unstable and ejected a part of its mantle to form the Moon (Cuk and Stewart, 2012). Researchers have started making further advancements and simulations based on impact hypothesis. Pahlevan and Stevenson (2007) simulated planet formation and concluded that the moon-forming impactor and the Earth did not have identical source. But, the model simulations do not account for the astounding similarity of oxygen isotopes, and other primordial isotopic compositions between the Earth and the Moon (Halliday, 2012). Recently, Wise (2014) has refuted the giant impact hypothesis stating it as unnecessary complication, but did not provide any new idea.

What are the issues that lead to the failure of the giant impact hypothesis? The objective of the present paper is to explain the formation process of the Moon in a new perspective.

## 2. Problems with the giant impact hypothesis

The geochemical compositions of the Earth and the Moon are found to be exceptionally similar (Elkins-Tanton, 2013). This should not have been affirmative if another planet having a different chemical composition had collided into the proto-Earth. If the impactor had a different origin, one must find its signatures in the samples from the Moon. This does not support the observations, however.

463, Wallace Building, Centre for Earth Observation Science, Department of Environment and Geography, Clayton H. Riddell Faculty of Environment, Earth, and Resources, University of Manitoba, Winnipeg R3T 2N2, Manitoba, Canada. [*]E-mail: guptm@yahoo.com. The author declares no conflict of interests. Submitted to arXiv on January 31, 2014.
© Mukesh Gupta 2014

If a giant planet collided with the Earth head-on, why did the newly formed giant object not deviate from its orbit around the Sun, it should have either left the Sun's orbit or formed a new orbit? The matter, which spread out into space (supposedly the matter which formed the Moon) because of the impact should have been travelling in a linear direction from the point of collision instead of the entire system circling around the Sun. Why did the orbit of other planet that collided into the proto-Earth, interfere with that of the Earth? It is logical that all the planets must have been in independent orbits during their formation processes. If the formation of the planets were not complete at the time of giant impact, where did the giant planet, the size of the proto-Earth, originate? On the contrary, if the formation process of all the planets in the Solar system were complete, all the planets must have been in their own orbits without any possibility of collision with another planet.

This makes one conclude that the formation process of most of the planets in the Solar system was not complete at the time of giant impact. But then, another question arises, how did the moons of other planets in the same Solar system form? This implies that there is a missing physical accord of the actual formation processes of planets and their moons. The giant impact hypothesis fails to explain the orbital dynamics, the disparities in tilt, and rotation axes of the Earth and the Moon. The physics of planet formation in the Solar system and the extrasolar systems must follow the same laws everywhere. It is posited here that the moons and planets have an analogous physical process of their formation.

Another proposition, called 'half-Earth impactor' is that the Moon formed resulting from collision of two planets, each of about half-Earth mass (Cuk and Stewart, 2012). This hypothesis is also facing a criticism and suffers from shortcomings that single impact hypothesis has. Where do the already-formed planets originate assuming the entire Solar system formed simultaneously, which is also evidenced from the ages of various meteorites, planets, and asteroids? The double impact hypothesis rather opposes formation of other planets and the moons in the entire Solar system. The present elucidation asserts that an impact is not a part of the Moon formation, or any planet formation. However, impacts can modify the orbital and axial dynamics, once the Moon has formed. I attempt to propose that the origin of the Moon was not a rare event; it has evolved following the same set of processes of formation as that of other planets and the moons in the Solar system.

## 3. Physics of the Moon formation

It is imperative to think that the formation of the Moon was mainly physical not chemical. The Moon did not come into existence chemically; instead, chemical processes started playing the role after the Moon had formed physically. Though, the chemical reorganization of matter and settlement into the present day form was a long process (~billions of years), but the first process of formation held deep physical concepts.

Formation of the Moon started from the gaseous cloud, which was present in the proto-Solar system (there was only gaseous cloud in a region of Milky Way before the formation of the Solar system). The Moon started forming simultaneously as other planets and their moons started forming. It is more convincing to use the term 'evolved' than 'formed' because this process did not occur instantaneously. The planets and the moons evolved through a seamless evolution process and all of them are still evolving on a time scale almost imperceptible at human scale. Where did the gaseous cloud come from? This question is generic in cosmology; however, gaseous cloud is the primitive matter that evolved into observed planetary bodies. This paper does not intend to discuss how and what happened before gaseous clouds.

In the Milky Way, the hot gaseous clouds revolve around the centre of the galaxy. A large portion of the gaseous matter resulted into a giant fireball because of accretion of matter as clumps, which grew over time, forming primitive Sun with very high gravity. Various regions/quantities of hot gaseous clouds were around the primitive Sun still not orbiting the Sun. As the Sun acquired enormous gravity compared to surrounding giant gaseous clouds, the region of primitive Sun's influence of gravity established. The gas dynamics, turbulence, irregular collisions, and abrupt motions of dense and light gaseous matter started, which resulted into matter either settling in separate giant fireballs near the primitive Sun or colliding into each other, with some matter flown away outside the primitive Solar influence. These processes occurred at a time scale of billions of years.

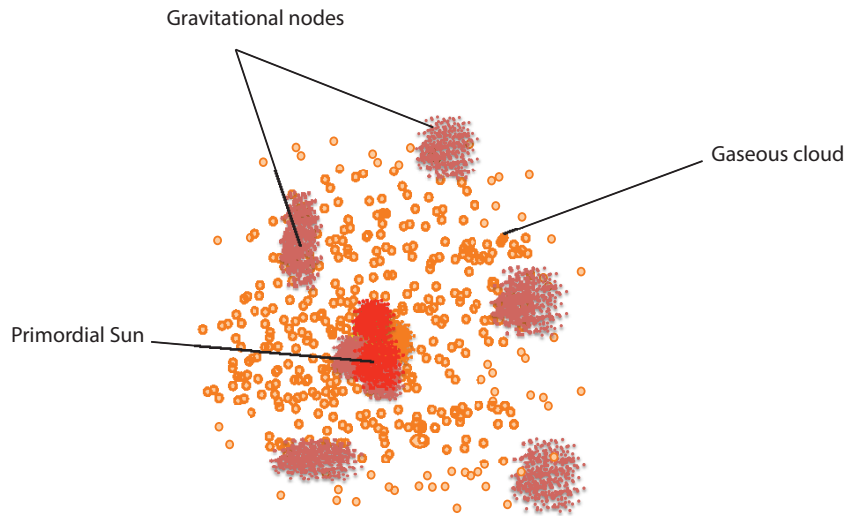

FIGURE 1. A part of gaseous disk viewed from above (not to scale). The gravitational nodal points are the regions of dense and higher gravity compared to surroundings. A large part of matter accumulated in the centre as primordial Sun. There were at least eight gravitational nodes, which ultimately evolved into planets that we see today. The Earth was one such gravitational node that had dust revolving around it.

Thus, the primitive Sun formed and the hot gaseous matter present in and around the gravitational influence of the Sun; set the grounds for the Solar system formation (Figure 1). There were still no orbits, no planets, and no moons, except the gaseous matter in dense pockets around the Sun. Over billions of years, the intense turbulence, haphazard giant collisions continued between a mix of gaseous and liquid matter as the hot gases started to cool. The gravitational nodal points (the region within the influence of the Sun where gravity of gaseous cloud was greater compared to other regions. These nodal points could have been more than eight – defined planets – but it is likely that some big planets perished in the collisions or expelled out of Sun's influence) around the Sun started forming as the matter cooled. Once these nodal points formed where gravity was higher than the surroundings, any further giant collisions into the gaseous/liquid fireballs resulted in their motion around the Sun because of angular momentum having imparted by collisions. Thus, the planets began taking shapes and set into orbital motion around the Sun. The gravitational forces of various gravity nodal points (planets) dominated the settlement of matter. Smaller gaseous fireballs came into gravitational influence of their parent planet depending on the size and gravity of a body. The distribution of matter in the gravitational influence zone of the Sun started stabilizing and empty space left behind as the matter either ended into a collision with the Sun/planet or coalesced into a planet/satellite. The debris that neither collided with the Sun nor with planets, kept travelling out of the Solar system or resulted into the unclassified inter-planetary objects such as Saturn rings, Van Allen belt, asteroids, etc. More and more collisions determined - the speed of revolution of a planet around the Sun, spin speed, tilt of axis, gravitational pull, and the distance between the bodies as the temperature subsided. The planets and their moons had come into existence; further collisions were not powerful enough to perturb the established orbital trajectories. The collisions at present are negligible that could dramatically change the physical equilibrium of the Solar system. The angular momentum was well balanced between planets and the moons, and between planets and the Sun. As the planetary matter (planets, moons, and the Sun) came into a physical equilibrium, chemical processes gained dominance as further cooling happened. Finally, at a later stage on geological time scale, the biological processes came into existence on the Earth as the chemical processes acquired the required equilibrium. Why biology evolved only on the Earth, is a separate topic of research and lies outside the scope of this paper.

Why did the Earth not have more than one Moon? The size of the Earth, its distance from the Sun and other planets, gravity, and general equilibrium required to sustain the Earth in its present form could be the reason only one Moon remained in its orbit. The argument that there could have been smaller hot gaseous clouds orbiting the proto-Earth can be possible; however, to obtain the physical equilibrium, the hot matter either collided with the proto-Earth and coalesced into it or set into a linear trajectory into space because of extra collisions and gravitational imbalance in the primitive Solar system.

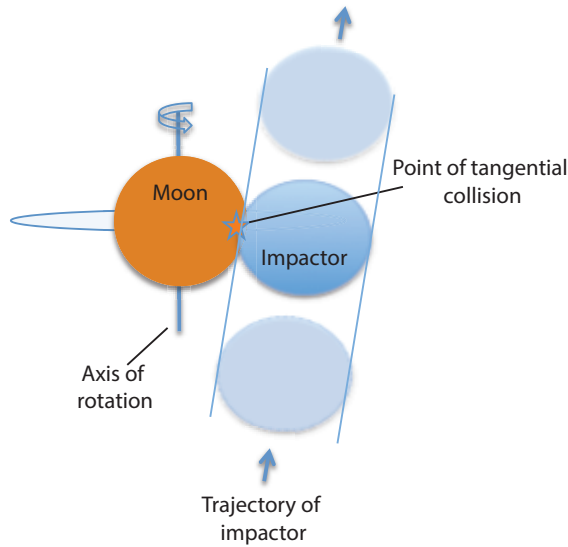

FIGURE 2. Cartoon showing what made the Moon (or Earth/other planets) rotate around its own axis. An impactor (a body of size comparable to the Moon) collides with the Moon tangentially brushing its surface to impart sufficient angular momentum. Other smaller impacts by asteroids, other terrestrial bodies, etc. partially modified its trajectory around the Earth, wobbling, precession, and rotation on its own axis. Analogous bombardments make all planetary bodies rotate around their own axis, and revolve around the Sun or a planet.

What kept the proto-Moon orbit the proto-Earth instead of colliding into the Earth? Both the Earth and the Moon evolved from the same gaseous cloud; one clump of cloud had higher gravity (greater mass) than the other. The physics of why and how the Earth started orbiting the Sun is the unchanged here also. First, a (or multiple) tangential impact (s) of a terrestrial body brushing the Earth's near surface imparting it sufficient tangential velocity to revolve in an elliptical orbit around the Sun and spin around its own axis, set off the Earth's revolution around the Sun (Figure 2). The spin might have modified by other smaller impacts giving the result that we observe today. Once the Earth-Moon system started orbiting the Sun in a physical equilibrium, an analogous but smaller tangential impact on the proto-Moon set off its revolution around the Earth and spin around its own axis (Figure 3). Other smaller impacts might have also modified the Moon's spin analogous to the Earth's spin. Note that not all tangential impacts would have resulted in setting off the revolution in a circular/elliptical orbit. The size, gravity of the body, and its distance from the object of revolution determine the elliptical or circular orbits. This is same as modern era artificial satellites get tangential velocity at apogee/perigee of the orbit to set off the revolution and inject the satellite into orbit around the Earth. This tangential velocity depends on the mass of the satellite, its distance from the Earth, and gravitational force of the Earth. Even a small error in the imparted tangential velocity may displace the satellite from the intended orbit, which could result into losing the satellite into deeper space. This analogy is also consistent for the natural satellite of the Earth. Because the proto-Moon comparatively was a giant spherical body, the imparted tangential velocity because of a collision also resulted in tilting its axis, and spinning.

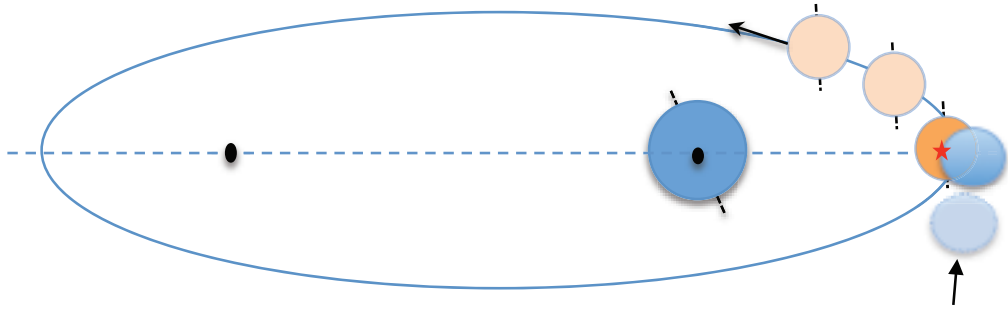

FIGURE 3. A collision at perigee (it can also be apogee) sets off the Moon to go around the Earth in an orbit (exaggerated eccentricity shown) that gets modified over billions of years because of subsequent bombardments of terrestrial bodies of comparable sizes (The Earth lies at one of the foci of the elliptical orbit). The Moon would have been stationary relative to the Earth at formation. It is the same principle that injects an artificial satellite into an orbit around the Earth. Analogous collisions in the Solar system created different orbits for different planets and their moons.

The Earth-Moon system had formed, and the cooling of the system continued. A thin crust of solid cooler magma formed on both the bodies. The evolution (cooling) rate of various planetary bodies and the moons were different due to their size, contents, composition, and distance from the Sun. The Moon cooled earlier and faster than the Earth because of its smaller size. The impacts by asteroids, meteorites, comets, and other wandering celestial objects continue till today (now at a much smaller rate). These impacts, after billions of years of stabilization of the Solar system, are now less frequent and not catastrophic to perturb the already established physical equilibrium of the Earth-Moon system. The deep craters left an indelible mark on the Moon surface because it did not have plate tectonics, crustal formation and subduction zones analogous to the Earth because of lack of convection currents in its mantle. Whereas, only about 11 km of the Earth's crust cooled, the remaining Earth is still a giant fireball consisting of convection currents of hot magma that causes geomagnetic field, plate tectonics, volcanoes, etc.

## 4. Implications

This perspective provides insights into how not only the Moon formed, but also how other planets and their moons formed. This elucidates why and how Uranus has its axis of rotation tilted to the plane of revolution around the Sun. As stated, there were collisions of all intensities from all directions in the primordial Solar system. The tangential impact that set off Uranus into an orbit around the Sun was uniquely different from the tangential impact that caused rotation around its own axis (Figure 3).

This also construes why Venus revolves in a retrograde direction around the Sun unlike other planets. When viewed from top of Sun's North Pole, all planets revolve anti-clockwise whereas Venus revolves clockwise. This resulted from the formation mechanism of Venus. Like the formation process of all the planets, the proto-Venus was also a gaseous cloud (a gravitational node) in the primordial Solar system. When matter started settling and cooling in the Solar system, collisions increased in the entire Solar system. Venus, as a giant fireball, also went through intense bombardment. But, a significant collision that set off the revolution of Venus around the Sun and rotation around its own axis happened to be from a different direction (Figure 3). It is noteworthy here that collisions were happening from all directions and were of all intensities. A question may arise here why Venus, the only planet among others, has a retrograde orbit? A tractable answer would be 'by probability' because collisions were occurring randomly from random directions. Not all collisions can set off the revolution of a big planet around the Sun. Irrespective of the direction of the giant tangential impact; only two possibilities of revolution exist, clockwise or anti-clockwise.

This perspective supports the observations of like geochemical compositions of the Earth-Moon system. Although, the rest of the Solar system originated from the same primordial gaseous cloud, but the size (gravity), rate of cooling, influence of the Solar radiation, distance from the Sun all mattered into what constituted in their chemical/isotopic compositions. The Earth and the Moon both went through similar processes of cooling because of proximity to each other, and had about same distance from the Sun on planetary scale. The Earth and the Moon formed from the same cloud of matter (the gravitational nodal point in the gaseous primordial Solar system).

## 5. Conclusions

The Moon evolved from the same gaseous cloud that the Earth originated from. Both the proto-Earth and the proto-Moon, as a system, started revolving around the Sun resulting from a tangential impact on the proto-Earth. This impact placed the Earth in an orbit around the Sun. An analogous but a smaller impact placed the Moon in orbit around the Earth. The rotation of the Moon on its own axis originated from tangential collisions to conserve the angular momentum. Angular momentum of the Earth and the Moon played a key role in the formation. The proto-Earth in gaseous cloud form would have been non-revolving, and a non-rotating body; and the Moon would have been stationary, hanging in one place at a distance from the Earth without the angular momentum, a collision would have resulted because of gravitational distinction between the two. If the giant impact hypothesis were correct, new moons should still form resulting from impacts.

A new perspective in this paper concludes that the moons can only form at the time of planet formation in a parallel and simultaneous process, and the new moons cannot form in a grown up Solar system. The Earth-Moon system was not special exception in nature; rather it formed following the same rules as other planets. The physics of scattering of matter because of an impact and computer simulations disapproves the giant impact hypothesis. It fails to interpret the scattered matter how it got recaptured into the Earth's orbit. The new perspective successfully brings insights into what happened between the disk formation and the accumulation of the Moon from the disk. If found conforming to reality, the perspective is likely to provide an impetus, fill the gaps, and solve many riddles that have puzzled scientists for long.